\documentclass[a4paper,fleqn,usenatbib,useAMS]{mnras}

\usepackage{graphicx}   
\usepackage{amsmath}    
\usepackage{amssymb}    
\usepackage{multicol}   
\usepackage{bm}         
\usepackage{pdflscape}  

\def\la{\;
\raise0.3ex\hbox{$<$\kern-0.75em\raise-1.1ex\hbox{$\sim$}}\; }
\def\ga{\;
\raise0.3ex\hbox{$>$\kern-0.75em\raise-1.1ex\hbox{$\sim$}}\; }

\newcommand{\dmm}{$\Delta\mu/\mu$}

\newcommand{\dV}{$\Delta V$}

\newcommand{\kms}{km~s$^{-1}$}
\newcommand{\ms}{m~s$^{-1}$}
\newcommand{\etal}{{et al.}}

\usepackage[T1]{fontenc}
\usepackage{ae,aecompl}

\title[Limits on $\mu$-variation in the Galaxy]
{\textit{
Probing the Electron-to-Proton Mass Ratio Gradient in the Milky Way
with Class I Methanol Masers
}}

\author[S. A. Levshakov \etal ] {
S. A. Levshakov$^{1,2,3}$\thanks{E-mail: lev@astro.ioffe.ru},
I. I. Agafonova$^{3}$,
C. Henkel$^{4,5,6}$,
Kee-Tae Kim$^{7,8}$,
\newauthor
M. G. Kozlov$^{2,3}$,
B. Lankhaar$^{9}$,
W. Yang$^{4}$
\vspace*{8pt}
\\
$^{1}$Ioffe Physical-Technical Institute, 194021 St.~Petersburg, Russia\\
$^{2}$Petersburg Nuclear Physics Institute of NRC "Kurchatov Institute", Gatchina, Leningrad District, 188300, Russia\\
$^{3}$Electrotechnical University ``LETI'', 197376 St.~Petersburg, Russia\\
$^{4}$Max Planck Institut f\"ur Radioastronomie, Auf dem H\"ugel 69, 53121 Bonn, Germany\\
$^{5}$Astron. Dept., King Abdulaziz University, PO Box 80203, 21589 Jeddah, Saudi Arabia\\
$^{6}$Xinjiang Astronomical Observatory, Chinese Academy of Sciences, 830011 Urumqi, People's Republic of China\\
$^{7}$Korea Astronomy and Space Science Institute, 776 Daedeokdae-ro, Yuseong-gu, Daejeon 34055, Republic of Korea\\
$^{8}$University of Science and Technology, Korea (UST), 217 Gajeong-ro, Yuseong-gu, Daejeon 34113, Republic of Korea\\
$^{9}$Department of Space, Earth and Environment, Onsala Space Observatory,
Chalmers University of Technology, Onsala, Sweden\\
}
\date{Accepted  Received ; in original form 2021 August }

\pubyear{2021}

\begin{document}
\label{firstpage}
\pagerange{\pageref{firstpage}--\pageref{lastpage}}
\maketitle

\begin{abstract}
We estimate limits on non-universal coupling of hypothetical hidden fields to standard matter
by evaluating the fractional changes in the electron-to-proton mass ratio, $\mu = m_{\rm e}/m_{\rm p}$,
based on observations of Class~I methanol masers distributed in the Milky Way disk over the range of 
the galactocentric distances $4 \la R \la 12$ kpc. The velocity offsets $\Delta V = V_{44} - V_{95}$ 
measured between the 44 and 95 GHz methanol lines provide, so far, one of the most stringent constraints 
on the spatial gradient $k_\mu \equiv d(\Delta \mu/\mu)/dR < 2\times10^{-9}$ kpc$^{-1}$
and the upper limit on \dmm~$< 2\times10^{-8}$, where
\dmm~= $(\mu_{\rm\scriptscriptstyle obs}-\mu_{\rm\scriptscriptstyle lab})/\mu_{\rm\scriptscriptstyle lab}$.
We also find that the offsets \dV\ are clustered into two groups which are separated by 
$\delta_{\scriptscriptstyle \Delta V} = 0.022 \pm 0.003$ \kms\ ($1\sigma$ C.L.).
The grouping is most probably due to the dominance of different hyperfine transitions in the 44 and 95 GHz methanol
maser emission. Which transition becomes favored is determined by an alignment (polarization) of the nuclear spins of
the four hydrogen atoms in the methanol molecule. This result confirms that there are
preferred hyperfine transitions involved in the methanol maser action.
\end{abstract}

\begin{keywords}
masers --
methods: observational --
techniques: spectroscopic --
ISM: molecules -- 
elementary particles  
\end{keywords}


\section{Introduction}
\label{Sec1}

A variety of theories for the dark sector (dark matter and dark energy)
suppose the existence of hidden fields which couple non-universally to
the Standard Model, SM
(for reviews, see Uzan 2011; Marsh 2016; Battaglieri \etal\ 2017; Hui \etal\ 2017;
Irastorza \& Redondo 2018; Terazawa 2018; Beacham \etal\ 2019; Ahmed \etal\ 2019).
Such coupling would modulate the fermion masses in different ways
thus changing their ratios. In particular, this relates to
the fundamental constant of particle physics~--
the electron-to-proton mass ratio, $\mu = m_{\rm e}/m_{\rm p}$,~-- where
$m_{\rm e}$ is directly determined by the coupling to the Higgs-like field(s),
whereas the main input to $m_{\rm p}$ comes from the binding energy of quarks.
Hence, measurements of $\mu$ can serve as a tool to probe the coupling between the SM and dark sector.
New fields are predicted to be ultralight (Compton wavelengths $\lambda \sim 1$~kpc)
and/or to change their values depending on the environmental parameters such as
gravitational potential of baryonic matter or local baryonic mass density
(e.g., Damour \& Polyakov 1994; Khoury \& Weltman 2004; Olive \& Pospelov 2008; Brax 2018).
This makes astronomical objects preferable targets in corresponding studies.

Measurements of fractional changes in $\mu$,
\begin{equation}
\frac{\Delta\mu}{\mu} = \frac{\mu_{\rm \scriptscriptstyle obs} -
\mu_{\rm \scriptscriptstyle lab}}{\mu_{\rm \scriptscriptstyle lab}},
\label{Eq1}
\end{equation}
in astronomical objects are based on the fact that the
molecular electron-vibro-rotational transitions have specific dependences on $\mu$
(Thompson 1975) and different sensitivities to
$\mu$-variations (Varshalovich \& Levshakov 1993; Flambaum \& Kozlov 2007; Levshakov \etal\ 2011;
Jansen \etal\ 2011; Patra \etal\ 2018).
For a given molecular transition frequency $f$,
the dimensionless sensitivity coefficient to a possible variation of $\mu$ is defined as\footnote{To avoid
confusion, we note that if
$\mu$ is defined as the proton-to-electron mass ratio, say $\mu' = m_{\rm p}/m_{\rm e}$, then
$\Delta\mu/\mu = -\Delta\mu'/\mu'$ and $Q_i = -K_i$, where $K_i = (df_i/f_i)/(d\mu'/\mu')$. }
\begin{equation}
Q = \frac{df/f}{d\mu/\mu}\, .
\label{Eq2}
\end{equation}
The coefficients $Q$ take positive or negative signs and values ranging between
$\sim 10^{-2}$ for H$_2$ and $\sim 10$s for
CH$_3$OH and other molecules (for a review see, e.g., Kozlov \& Levshakov 2013).
The fractional changes in $\mu$ can be measured using
any pair of lines of co-spatially distributed molecular transitions ($i, j$)
with different values of $Q$ (Levshakov \etal\ 2011):
\begin{equation}
\frac{\Delta\mu}{\mu} = \frac{V_j - V_i}{c(Q_i - Q_j)},
\label{Eq3}
\end{equation}
where $V_j$ and $V_i$ are the LSR radial velocities
of molecular transitions with sensitivity coefficients $Q_j$ and $Q_i$,
and $c$ is the speed of light.
It is to note that spectral observations with modern facilities
provide an unprecedented accuracy in measurements of molecular transitions
and, hence, in \dmm. Additional advantages are the relative simplicity of
interpretation of the obtained results
and a restricted number of sources of systematic
errors (cf., e.g., Touboul \etal\ 2020).

Presently, the most stringent limit on \dmm,
\dmm~$< 7\times10^{-9}$ (hereafter a $1\sigma$ confidence level is used),
was obtained from
high resolution spectral observations of Milky Way's cold
molecular cores in lines of NH$_3$, HC$_3$N, HC$_5$N, HC$_7$N, and N$_2$H$^+$
at the Effelsberg 100-m, Medicina 32-m, and Nobeyama 45-m radio telescopes
(Levshakov \etal\ 2010a,b; Levshakov \etal\ 2013, hereafter L13).
Additionally, observations of the dense dark cloud core L1498 
in thermal $E$- and $A$-type methanol CH$_3$OH lines at
the IRAM 30-m telescope gave the upper limit of \dmm~$< 2\times10^{-8}$
(Dapr\'a \etal\ 2017, hereafter D17).
In these studies, all molecular cores were located within a 300 pc radius from the Sun,
which is insufficient to detect the predicted gradients of hidden ultralight fields:
to probe the SM--dark sector coupling, 
observations of targets spaced apart by distances of kiloparsecs are required.

Such targets are objects of the present paper.
We aim at obtaining the \dmm\ estimate in the Milky Way disk utilizing
narrow emission lines of bright Class~I methanol (CH$_3$OH) masers
from the northern Galactic hemisphere distributed
over the galactocentric distance range of $4 \la R \la 12$ kpc.

Methanol masers are usually classified into two types: Class~I and Class~II.
The first type of sources is offset by $\sim1$ pc from star formation signposts
and mostly the result from collisional excitation
(Menten 1991a,b; Cragg \etal\ 2005; Leurini \etal\ 2016).
The sources are associated with shocks caused by molecular outflows, expansion of H\,{\sc ii} regions,
and cloud-cloud interactions (Kurtz \etal\ 2004; Voronkov \etal\ 2006, 2010, 2014).
Masers of the second type are found in the closest environment of massive young stellar
objects and are pumped by the reprocessed dust continuum radiation from these sources.
The Class~II methanol masers at 6.2 and 12.2 GHz were used 
by Ellingsen \etal\ (2011) to limit $\mu$-variations at the level 
$|\Delta\mu/\mu| < 2.7\times10^{-8}$. 
The advantages of using Class~I methanol masers for the \dmm\ measurements are the following:
\begin{itemize}
\item[$\bullet$]
They  belong to a large population of
Galactic emitters which are distributed
across the Galactic plane towards both the Galactic centre and anti-centre.
This enables us to scan \dmm\ over a large spatial range.
\item[$\bullet$]
The maser lines are strong and narrow (non-thermal) and, thus, their radial velocities
can be measured with high precision.
\item[$\bullet$]
Methanol transitions show large differences in the sensitivity coefficients
(denominator in Eq.~\ref{Eq3}). This naturally decreases
uncertainties of the \dmm\ estimate.
\item[$\bullet$]
Class~I masers are stable, i.e., do not exhibit 
flux variability at time intervals $\sim 1$~yr
in contrast to Class~II methanol masers which show greater temporal variability.
\end{itemize}

We estimate the fractional changes in $\mu$ using
the Class~I $A$-type methanol transitions at 44 and 95 GHz.
The emission in these transitions is closely associated and traces the same spots
(Val'tts \etal\ 2000; Voronkov \etal\ 2014; Leurini \etal\ 2016).
This minimizes the Doppler noise, which are random shifts of spectral line positions
caused by possible spatial segregation and kinematic effects.
In addition to this component, which is stochastic, the total error budget of the measured radial velocity,
$V_{\scriptscriptstyle\rm LSR}$,
of a given spectral line contains also a systematic error related to the uncertainty in the
laboratory values of rest frequencies.
The laboratory frequencies of the 44 and 95 GHz transitions are  presently measured with an error of
$\sim 10$ kHz (Tsunekawa \etal\ 1995; M\"uller \etal\ 2004).
This translates into systematic errors of $V_{\scriptscriptstyle\rm LSR}$
as great as $\sim 10$ \ms, restricting the limit on \dmm\ at the level of $10^{-8}$,
i.e., the level already reached in the most accurate \dmm\ estimates up-to-now.

In principle, it is technically possible to measure the laboratory
frequencies with accuracy $\sim 1$ kHz. However, the problem is more complicated.
Namely, the molecule CH$_3$OH possesses an underlying hyperfine structure
scaled over the same $\sim 10$ kHz
(Hougen \etal\ 1991; Coudert \etal\ 2015; Belov \etal\ 2016; Lankhaar \etal\ 2016).
The hyperfine splitting can only be partly resolved in the laboratory and
is completely convolved in astrophysical observations.
If emission is purely thermal, then the barycentre of the convolved profile
is more or less stable and its error is localized within the kHz uncertainty interval (D17).
If, however, emission is due to masing effects, then, as shown in
Lankhaar \etal\ (2018), population inversion may be enhanced for some hyperfine transitions, 
while suppressed for others. The physical mechanisms leading to the favoured pumping of this 
or that hyperfine sublevel are not yet clear enough. As a result, the barycentre of the convolved 
profile will shift with an amplitude of $\sim 10$ kHz depending on the physical environment.
The investigation of this additional systematics is another purpose of the present paper.

\section{Observations and target selection}
\label{Sec2}

We use simultaneous observations of the 44 GHz ($7_0 - 6_1$~A$^+$) and 95 GHz ($8_0 - 7_1$~A$^+$)
Class I methanol masers from the so-called Red {\it MSX} Source (RMS) 
catalogue\footnote{https://rms.leeds.ac.uk/cgi-bin/public/RMS\_DATABASE.cgi}
observed by Kim \etal\ (2018) (hereafter K18), and
from the Bolocam Galactic Plane Survey (BGPS) 
sources\footnote{https://irsa.opac.caltech.edu/data/BOLOCAM\_GPS}
observed by Yang \etal\ (2020) (hereafter Y20).
Both surveys were performed with the Korean VLBI Network (KVN) 21-m telescopes in single-dish telescope mode.
The multifrequency receiving systems at each telescope allow us to observe 44 and 95 GHz methanol transitions
at the same time, and with velocity scales defined in the same way. These two advantages are not being
met by any other data set. Details of these observations are given in the original papers,
here we repeat only those relevant to the present study.

\begin{figure*}
\vspace{-3.0cm}
\includegraphics[angle=0,width=1.0\linewidth]{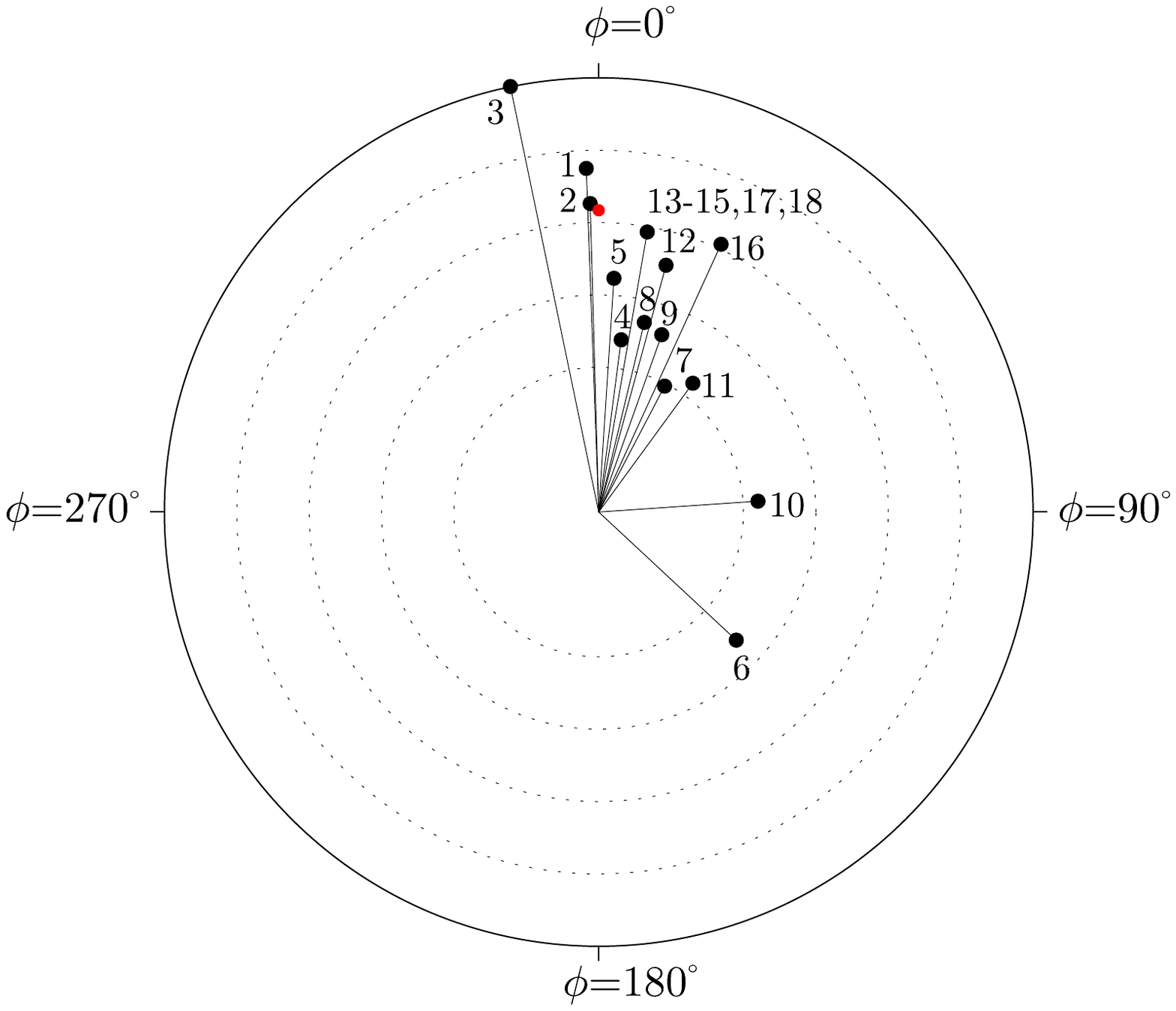}
\vspace{-5.0cm}
\caption{
Location of the methanol masers
(marked by numbers in accord with Table~\ref{Tb1})
in projection onto the Galactic plane.
The concentric circles give
the galactocentric distances starting from 4 kpc; the increment is 2 kpc.
The Galactic centre is at coordinates $(R,\phi) = (0,0)$, and the Sun (red dot) is
at coordinates $(R,\phi)$ = (8.34~kpc,0), as adopted by M\`ege \etal\ (2021).
}
\label{fg1}
\end{figure*}

Both the K18 and Y20 works were surveys aimed in the  first place
at the discovery of new methanol masers.
Because of that the signal-to-noise ratio (S/N) in the majority of targets
was not high enough to deduce the radial velocities with the required accuracy
of $\sim$~10s \ms. As a result, only 7 objects from 229 maser sources of K18 and
11 objects from 144 sources of Y20
turned out to be suitable for our purposes; the selected targets
are listed in Table~\ref{Tb1}.
Their spatial distribution in the Galactic disk is shown in Fig.~\ref{fg1}.
The objects lie at low Galactic latitudes $-2.6^\circ < b < 2.8^\circ$ and
are located towards both the Galactic centre and anti-centre at distances
$D_a \leq 4.7$ kpc (anticentre) and $D_c \leq 12.3$ kpc (centre) from the Sun.
Thus, the maximum scale covered is approximately 16 kpc which is
comparable to the Galactic optical radius.

Both the K18 and Y20 surveys report a remarkable strong correlation between methanol 
peak velocities for the 44 and 95 GHz transitions:
\begin{equation}
V_{\rm pk,95} = (0.98\pm0.05)V_{\rm pk,44} - (0.01 \pm 0.010),
\label{Kim}
\end{equation}
and
\begin{equation}
V_{\rm pk,95} = (0.9998\pm0.0010)V_{\rm pk,44} - (0.0470 \pm 0.0614),
\label{Yong}
\end{equation}
respectively.
For the mean value of ${V}_{\rm pk,44} \simeq 75$ \kms\ (see Fig.~4 in Y20)
the relation (\ref{Yong}) provides for the offset $\Delta V = V_{\rm pk,44} - V_{\rm pk,95}$
an error $\sigma_{\scriptscriptstyle \Delta V} \sim 0.1$ \kms, which transforms into an upper
limit on \dmm~$< 10^{-7}$ in accord with Eq.(\ref{Eq3}) where $Q_{95} - Q_{44} = 3.3$ 
(see Subsec.~\ref{SSec33}). Below we show how a more stringent limit on \dmm\  can be deduced from
the existent datasets. 

In K18, the original channel widths were $\Delta_{\rm ch} = 0.053$ \kms\
at 44 GHz, and 0.025 \kms\ at 95 GHz, whereas in Y20, they were two times larger,
$\Delta_{\rm ch} = 0.106$ \kms\ at 44 GHz, and 0.049 \kms\ at 95 GHz.
The peak flux densities, $F_{44}$ and $F_{95}$, indicated in Table~\ref{Tb1},
vary between 12 Jy (95 GHz, RMS153) and 420 Jy (44 GHz, BGPS4252) providing
signal-to-noise ratios between S/N = 10 and S/N = 276.
Both the signal-to-noise ratio and channel width are crucial for our analysis since they define
the final uncertainty of the line position. Namely,
an expected statistical uncertainty of the centre of a Gaussian-like
line profile is given by (e.g., Landman \etal\ 1982):
\begin{equation}
\sigma_0 \simeq \frac{0.7 \Delta_{\rm ch}}{\rm S/N} \sqrt{n},
\label{Eq4}
\end{equation}
where $n = {\rm FWHM}/\Delta_{\rm ch}$ is the line's full width at half maximum (FWHM) in units of channels.
With a typical line width (FWHM) for the methanol lines in question of $\sim 0.5$ \kms, one obtains
for the marginal values of S/N = 10 and 276 the following errors:
$\sigma_0 \simeq 0.014$ \kms\ ($\Delta_{\rm ch} = 0.074$ \kms) and
$\sigma_0 \simeq 0.0006$ \kms\ ($\Delta_{\rm ch} = 0.106$ \kms).

The adopted rest frequencies 44069.430 MHz ($7_0 - 6_1$~A$^+$) from Pickett \etal\ (1998)
and 95169.463 MHz ($8_0 - 7_1$~A$^+$) from M\"uller \etal\ (2004)
were utilized in the original papers of K18 and Y20.
Note that these frequencies differ from those recommended by NIST, NRAO, JPL, and CDMS
(see Table~\ref{Tb2}).

\section{Analysis}
\label{Sec3}

\subsection{Method}
\label{SSec31}

At first, we determined the baseline for each original spectrum
by choosing spectral windows without emission lines and/or noise
spikes and then calculating the mean flux densities $F_i$
along with their rms uncertainties $\sigma_i$ for each
spectral window.  Using spline interpolation through this set of pairs $\{F_i,\sigma_i\}$
we calculated a baseline which was subtracted from the spectrum.
Then, the mean value of the rms uncertainties, $\sigma_{\rm rms}$, was determined
and assigned to the whole spectrum.

The radial velocities $V_{\scriptscriptstyle\rm LSR}$ in Eq.~\ref{Eq3}
are calculated as described in Levshakov \etal\ (2019).
$V_{\scriptscriptstyle\rm LSR}$ is attributed to the line centre which is defined as a point 
where the first order derivative of the line profile is equal to zero. In order to calculate this 
extremum point accurately the observed line profile, $f(x)$, is filtered. 
The filtering function $y(x)$ consists of a sum of $N$ Gaussian subcomponents,
parameters of which are calculated by minimization of a $\chi^2$ function:
\begin{equation}
\chi^2_\nu = \frac{1}{\nu}\sum \{[f(x_i) - y(x_i)]/\sigma_{\rm rms}\}^2,
\label{Eq4a}
\end{equation}
where $\nu$ is the number of degrees of freedom.
The number of subcomponents, $N$, is chosen so that the $\chi^2_\nu$ function
is minimized at the level of $\chi^2_\nu \simeq 1$ to avoid under- or
over-fitting of the line profile. 
The uncertainty of $V_{\scriptscriptstyle\rm LSR}$, $\sigma_v$, is determined by
three points $\{x_1,y_1; x_2,y_2; x_3,y_3\}$ with $x_1 < x_2 < x_3$
which include the flux density peak, $x_{\rm peak} \in (x_1,x_3)$:
\begin{equation}
\sigma_v = \frac{\sigma_{\rm rms}\cdot \Delta_{\rm ch}}{(y_1 - 2y_2 + y_3)^2}{\cal K},
\label{Eq5}
\end{equation}
where ${\cal K} =  \sqrt{(y_3-y_2)^2 + (y_1-y_3)^2 + (y_2-y_1)^2}$, and
the channel width $\Delta_{\rm ch} = x_2-x_1 = x_3-x_2$.

\subsection{Reproducibility of radial velocities}
\label{SSec32}

The 44 GHz transition was observed by K18 in two epochs:
the first one in 2011, and the second in 2012.
The second set of observations included also the 95 GHz line.
The one year time lapse can be used to test stability
of Class~I methanol masers as was mentioned in Sect.~1.

\begin{figure*}
\includegraphics[angle=0,width=1.0\linewidth]{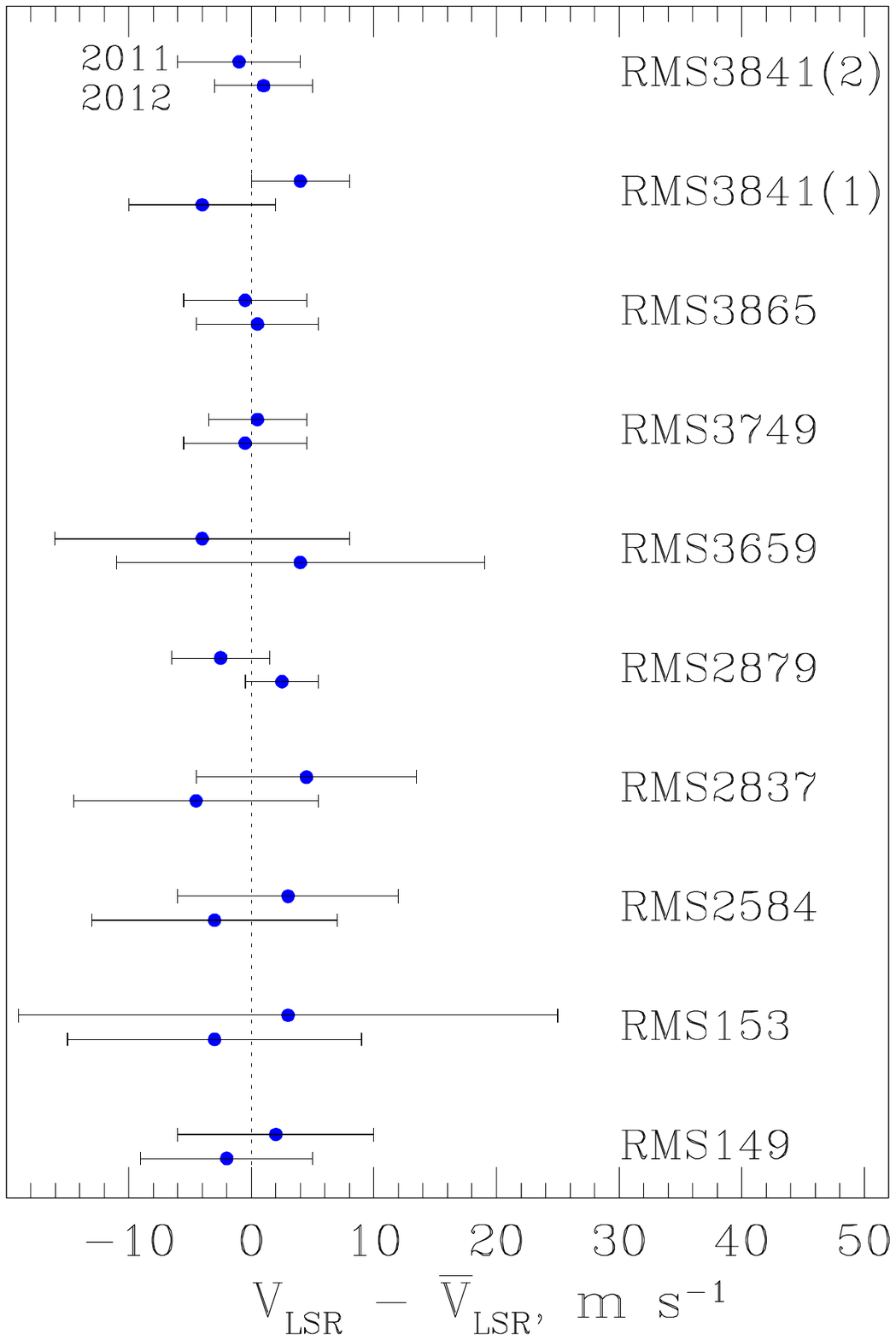}
\vspace{-4.5cm}
\caption{Reproducibility of the methanol line positions illustrated by the
relative offsets of the measured $V_{\rm \scriptscriptstyle LSR}$ velocities of
the 44 GHz line with respect to the mean value $\bar{V}_{\rm \scriptscriptstyle LSR}$
between the two observational epochs indicated at the left hand upper corner
(see Table~\ref{Tb3} for details).
Error bars represent the experimental uncertainty $(1\sigma)$.
The weighted mean of the velocity offsets
$\Delta V_{\rm yr} = V_{2011} - V_{2012}$ is $\langle \Delta V_{\rm yr} \rangle_w = -1.1\pm1.6$ \ms.
}
\label{fg2}
\end{figure*}

The radial velocities $V_{\rm \scriptscriptstyle LSR}$
of the 44 GHz line measured with respect to the mean value $\bar{V}_{\rm \scriptscriptstyle LSR}$
between the two observational epochs are plotted in Fig.~\ref{fg2},
while the individual $V_{\rm\scriptscriptstyle LSR}$ values are listed in Table~\ref{Tb3}.
This table includes 10 velocity offsets between 2011 and 2012 observations towards 9 maser sources
(the 44 GHz profile in RMS3841 consists of two narrow subcomponents separated by 0.4 \kms).
However, in the further analysis we used only 7 targets since the 95 GHz profiles towards
RMS2584 and RMS3841 were not good enough for precision measurements of their 
$V_{\rm \scriptscriptstyle LSR}$ values.

Figure~\ref{fg2} demonstrates stability of the 44 GHz line position for all 10 pairs with
the weighted mean $\langle \Delta V_{\rm yr} \rangle_w = -1.1\pm1.6$ \ms.
Taking this into account, we stack up 44 GHz spectra from both epochs
coadding them with weights inversionally proportional to their
variances, $\sigma^2_{\rm rms}$.
The measured radial velocities based on the stacked data are
listed in Table~\ref{T4}, second column.
Since for the 44 and 95 GHz data, being taken at the same time (in 2012) with the same telescopes, 
the computations of velocity corrections leading to LSR velocities match each other. Thus
stacking does not introduce a statistically significant systematic error.

\subsection{Fractional changes in $\mu$ }
\label{SSec33}

As it follows from Eq.~\ref{Eq3}, the fractional changes in $\mu$ are defined by
the LSR radial velocities of a pair of methanol lines, $V_i$ and $V_j$, which have
different sensitivity coefficients, $Q_i$ and $Q_j$, to $\mu$-variations.

For the 44 and 95 GHz methanol transitions the sensitivity coefficients were calculated
in Jansen \etal\ (2011) and in Levshakov \etal\ (2011). Both groups give similar $Q$-values:
$Q_{44} = -5.2\pm0.3$,
$Q_{95} = -1.88\pm0.09$ (Jansen \etal), and
$Q_{44} = -5.3\pm0.6$ and
$Q_{95} = -1.9\pm0.3$ (Levshakov \etal).
In our analysis we use $Q_{44} = -5.2$ and $Q_{95} = -1.9$.
Their difference $\Delta Q = 3.3$ is comparable to the difference between the sensitivity
coefficients in the ammonia method where $\Delta Q = 3.46$
(Flambaum \& Kozlov 2007; Levshakov \etal\ 2010a).

\begin{figure*}
\includegraphics[angle=0,width=1.0\linewidth]{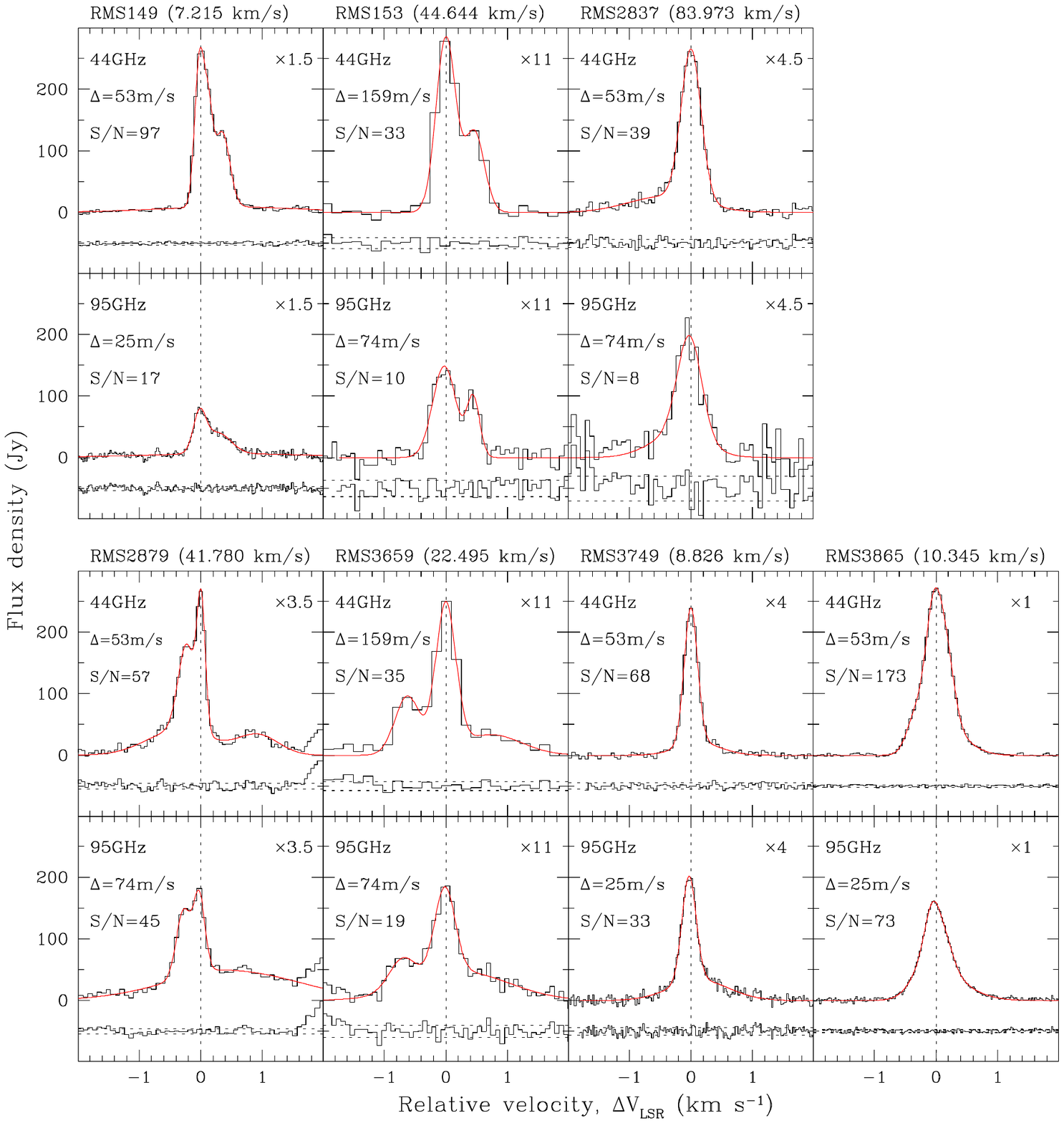}
\vspace{-4.5cm}
\caption{
Black histograms are the baseline subtracted emission lines of CH$_3$OH
at 44 GHz and 95 GHz towards
RMS sources (Red {\it Midcourse Space Experiment}) taken from Kim \etal\ (2018)
and listed in Table~\ref{T4}.
The LSR radial velocities are given relative to the peak LSR velocity of the 44 GHz line
which is indicated in parentheses after the source name.
The fitting curves are shown by red.
The residuals are plotted by the lower black histogram (arbitrarily
offset for clarity), while the horizontal dotted lines show their $\pm1\sigma$ boundaries.
The signal-to-noise ratio (S/N) per channel
at the line peak, the used channel width $\Delta$ (in \ms), and the flux density scale factor are
depicted in each panel.
The vertical dotted line marks the peak position of the 44 GHz line given to
indicate small velocity offsets for the 95 GHz line.
}
\label{fg3}
\end{figure*}

The measured LSR radial velocities along with their uncertainties
are given in Tables~\ref{T4} and \ref{T5}, whilst the fitting procedure is
illustrated in Figs.~\ref{fg3}~--~\ref{fg5}.
In these figures, vertical panels represent the 44 GHz (upper panel) and 95 GHz (lower panel) profiles of
methanol masers whose names and peak velocities of the 44 GHz line are indicated at the top of each block.
Some spectra were smoothed in order to improve the S/N ratio 
in individual channels and the used channel width is indicated in the corresponding panel.

In all calculations we used the rest frequencies adopted in the original papers of K18
and Y20 (see Table~\ref{Tb2}):
$f^{\rm\scriptscriptstyle P}_{44} = 44069.430$ GHz from Pickett \etal\ (1998), and
$f^{\rm\scriptscriptstyle M}_{95} = 95169.463$ GHz from M\"uller \etal\ (2004).
If other sets of the rest frequencies listed in Table~\ref{Tb2} would be adopted,
then the velocity offsets, $\Delta V = V_{44} - V_{95}$,
change as shown in Table~\ref{T4}, columns 4-8. These values were calculated in the following way.

The KVN telescopes adopt the radio definition of radial velocity which is given by
\begin{equation}
V_{\rm\scriptscriptstyle LSR} = c\left(1 - \frac{f_{\rm obs}}{f_{\rm lab}}\right),
\label{Eq6}
\end{equation}
where $c$ is the speed of light.
If we now consider two methanol lines with laboratory rest frequencies $f'_{0,i}$ and $f'_{0,j}$
(``old'' reference frame)
which are observed at the corresponding sky frequencies $f_i$ and $f_j$ and have the LSR radial
velocities $V'_i$ and $V'_j$, then the difference between the measured velocities of these lines
is determined for a new set of laboratory frequencies $f_{0,i}$ and $f_{0,j}$
(``new'' reference frame) as
\begin{equation}
\Delta V_{ij} = \Delta V'_{ij} + \delta V_{ij},
\label{Eq7}
\end{equation}
where $\Delta V'_{ij} = V'_i - V'_j$ is the velocity offset between the lines $i$ and $j$
in the ``old'' reference frame, and
$\delta V_{ij}$ is the Doppler correction term between the ``old'' and ``new'' reference frames:
\begin{equation}
\delta V_{ij} = c\left( \frac{f'_{0,j}}{f_{0,j}} - \frac{f'_{0,i}}{f_{0,i}} \right).
\label{Eq8}
\end{equation}

The absolute values of the weighted mean velocity offsets under different sets of laboratory frequencies
$| \langle \Delta V \rangle_w |$ range from 0.030 \kms\ to 0.176 \kms\ (Table~\ref{T4}).
In terms of \dmm\ (Eq.~\ref{Eq3}) these boundaries correspond to
\dmm~= $3.0\times10^{-8}$ and $1.8\times10^{-7}$, respectively. However, the latter clearly exceeds the
upper limits on \dmm\ found in the Galaxy: \dmm~$< 7\times10^{-9}$ (L13) and \dmm~$< 2\times10^{-8}$ (D17).
This means that the uncertainties in the JPL, NIST, and CDMS catalogues
as well as those reported in Tsunekawa \etal\ (1995) seem to be far too small
since the published rest frequencies lead to unrealistically large estimates of \dmm~$\sim 10^{-7}$.

Accounting for all these details, we list in Table~\ref{T5} only those
$V_{\rm\scriptscriptstyle LSR}$ and \dV\ values which
are calculated with rest frequencies taken from the original paper of Y20.

Our final sample consists of 7 sources from K18 and 11 sources from Y20
and the source BGPS6820 provides two peaks (see Fig.~\ref{fg5}),
i.e., in total we have 19 velocity offsets.

\subsection{Spatial gradient of \dmm}
\label{SSec34}

For each of the selected targets the measured velocity offset \dV~= $V_{44} - V_{95}$
is depicted in the upper panel of Fig.~\ref{fg6} against the target's distance $R$ from the Galactic centre.
The blue squares and red dots represent the sources from Tables~\ref{T4} and \ref{T5},
respectively. The indicated numbers correspond to the numbering in Table~\ref{Tb1}.
The source BGPS6820 is represented by two red dots $15_1$ and $15_2$.

It is seen that there is a clustering of points into two groups what hints to
the possible bimodality of the underlying distribution of the velocity offsets.
This hypothesis is statistically tested (by $\chi^2$-criterium) as illustrated in the lower panel
of Fig.~\ref{fg6}.
The unimodal distribution (black curve) with the mean $\langle \Delta V \rangle = 0.024$ \kms, dispersion
$\sigma = 0.012$ \kms, and the number of degrees of freedom $n = 7$ gives $\chi^2 = 9.9$,
which corresponds to a significance (probability) of 20\%.
On the other hand, the bimodal distribution (red curve) with $n = 5$,
two separate means $\langle \Delta V \rangle_1 = 0.032$ \kms,
$\langle \Delta V \rangle_2 = 0.010$ \kms, and dispersions $\sigma_1 = 0.006$ \kms,
$\sigma_2 = 0.004$ \kms\ delivers $\chi^2 = 3.9$ with the significance of 60\%.
Thus, we separate the data into two subsamples with $N = 12$ and $N = 7$ points.
The corresponding sample means and their $1\sigma$ errors are
$\langle \Delta V \rangle_1 \equiv \bar{x} = 0.0316\pm0.0018$ \kms, and
$\langle \Delta V \rangle_2 \equiv \bar{y} = 0.0104\pm0.0015$ \kms,
(marked by the horizontal dashed and dotted lines in the upper panel of Fig.~\ref{fg6}).
The weighted means and their errors are similar:
$\langle \Delta V \rangle_{1,w} = 0.0321\pm0.0014$ \kms, and
$\langle \Delta V \rangle_{2,w} = 0.0086\pm0.0009$ \kms.
The significance of this difference in terms of Student's $t$-test
with $n = 17$ degrees of freedom is about $7\sigma$:
$\delta = (\bar{x}-\bar{y}) \pm \sigma(\bar{x}-\bar{y}) = 0.022 \pm 0.003$ \kms.

Additional arguments in support of two groups
are that each of them contains points from both the K18 and Y20 surveys
and that there is no correlation with the galactocentric distances, $R$, and/or
the bolometric luminosities, $L_{\rm bol}$, listed in Table~\ref{Tb1}.
This minimizes the probability of possible observational selection and systematic biases.

We note that the revealed bimodality does not depend on the rest frequencies
of the 44 and 95 GHz lines
since any combination of the rest frequencies listed in Table~\ref{Tb2} would simply lead to
a parallel shift of all points in the upper panel of Fig.~\ref{fg6} along the $Y$-axis.
With our adopted set of the rest frequencies
the weighted means for \dmm\ for each group are the following: 
$(\Delta \mu/\mu)_{1,w} = (32.5\pm1.4)\times10^{-9}$
and $(\Delta \mu/\mu)_{2,w} = (8.6\pm0.9)\times10^{-9}$.

Table~\ref{Tb1} shows that the targets are distributed at galactocentric distances $4.0 \la R \la 12.3$ kpc.
In this range, the input of the baryonic gravitational potential to the circular velocity of the Galactic 
rotation falls from 60-70\% at $R \sim 4$ kpc to 30-50\% at $R \sim 12$ kpc, 
whereas the input attributed to the dark matter increases correspondingly 
(e.g., Eilers \etal\ 2019; Bobylev \etal\ 2021; Nitschai \etal\ 2021).
If the putative coupling between the baryonic and dark matter exists, then one can expect
some dependence of \dmm\ on $R$. 

The linear regression analysis of the velocity offsets for both groups returns
similar constraints on the gradient of \dV:
$(k_{\scriptscriptstyle \Delta V})_1 = -0.0015 \pm 0.0020$ km~s$^{-1}$~kpc$^{-1}$ in the
range $4.0 \la R \la 12.3$ kpc, and
$(k_{\scriptscriptstyle \Delta V})_2 = -0.0002 \pm 0.0027$ km~s$^{-1}$~kpc$^{-1}$ in the
range $4.5 \la R \la 8.7$ kpc,
what gives an upper limit on the gradient $k_\mu$ of \dmm\ in the Milky Way disk:
$k_\mu < 2\times10^{-9}$ kpc$^{-1}$.
With this gradient, we constrain the value of \dmm\
in the range $4.0 \la R \la 12.3$ kpc as \dmm~$< 2\times10^{-8}$.
This estimate is only slightly better than
that reported by D17 and Ellingsen \etal\ (2011), but it is three times less tight as in L13 due to
poor statistics of our targets.

It is interesting to compare the obtained $k_\mu$ estimate with corresponding values from other experiments.
For example, the MICROSCOPE satellite mission also aimed at measuring the effects of possible non-universal coupling,
but in a completely different way: it placed two masses of different composition (titanium
and platinum alloys) on an orbit around the Earth
and measured their circular acceleration (torsion balance). The reported upper limit on
the non-standard coupling is $< 10^{-14}$ (Touboul \etal\ 2020). 
With our estimate of the spatial gradient and the assumption of its isotropy one
obtains for the MICROSCOPE distance from the Earth centre (7000 km) an upper limit on the non-standard
coupling of $< 4\times10^{-22}$ which is tighter by eight orders of magnitude.
This shows once more the advantages of astrophysical spectroscopic methods.

\section{Maser action in the 44 and 95 GHz hyperfine components}
\label{Sec4}

The clustering of the measured velocity offsets \dV\ into two groups separated by $\delta = 22\pm3$ \ms\
raises a question about the physical mechanism behind the observed effect.
Since the total nuclear spin angular momentum, {\bf I}, of the molecule 
CH$_3$OH is non-zero for the torsion-rotation states,
each state of the symmetry A participating in the maser transition 
is split up by hyperfine interactions in a pattern
of $2I+1$ hyperfine states each of which is characterized by the total angular momentum
{\bf F} = {\bf J} + {\bf I}, with {\bf J} being the rotational angular momentum.
For transitions involving $A$-type levels with total nuclear spin $I = 1$ and 2,
the shift of 22 \ms\ ($\sim 3$ and 7 kHz respectively, at 44 and 95 GHz) is comparable
with the hyperfine frequency shifts within the corresponding torsion-rotation pattern.
The frequency shifts, $\Delta f$, between the strongest hyperfine components with $\Delta F = \Delta J = 1$
(Lankhaar \etal\ 2016) are listed in Tables~\ref{T6} and \ref{T7} for, respectively,
the 44 and 95 GHz transitions and are schematically illustrated in Figs.~\ref{fg7} and \ref{fg8}.

These data show that the $\approx 22$ \ms\ separation could occur if
the maser action is limited to a particular hyperfine
transition with the largest Einstein $A$-coefficient for spontaneous emission.
Namely, if the favoured transitions of the first group are
those with the smallest values of $F = J-2$,
$F = 5a \rightarrow 4a$ (44 GHz)
and $F = 6a \rightarrow 5a$ (95 GHz), whereas for the second group they are
those with the largest values of $F = J+2$,
$F = 9a \rightarrow 8a$ (44 GHz) and $F = 10a \rightarrow 9a$ (95 GHz),
then the velocity offset between the two groups would be 23 \ms.

The masing process with the dominance of particular hyperfine transitions
has already been suggested by Lankhaar \etal\ (2018)
for interpretations of the circular-polarization observations
of Class~II methanol masers at 6.7 GHz ($5_1 \rightarrow 6_0 A^+$).
Two favoured transitions $F = 3 \rightarrow 4$ ($F = J-2$)
and $F = 7 \rightarrow 8$ ($F = J+2$)
with the largest Einstein $B$-coefficient for stimulated emission\footnote{The $B$ coefficient for
stimulated emission from an upper level 2 to a lower level 1 is defined as
$B_{21} = (\pi^2 c^3/\hbar\omega^3)A_{21}$ (e.g., Hilborn 1982).}
were used to deduce an average magnetic field strength $\langle |B| \rangle \approx 12$ mG
in protostellar disks which was in line with OH maser polarization observations.
However, including all CH$_3$OH hyperfine components would lead to a considerably larger magnetic
field strength $\langle |B| \rangle \approx 80$ mG.

Similar maser actions limited to favoured hyperfine transitions
were considered by Lankhaar \etal\ (2018) in their study of polarization
observations of Class~I methanol masers at 36 and 44 GHz
in the outflows of massive star-forming regions.
For the $E$-type levels $4_{-1} \rightarrow 3_0$ at 36 GHz (total nuclear spin $I = 0$ and 1),
the $F = 3 \rightarrow 2$ hyperfine line with the smallest value of  $F = J-1$
was found to be dominating.
For the $A$-type levels at 44 GHz, one favoured transition was $F = 5a \rightarrow 4a$
as in our case, but for the other line there are two options with the largest
Einstein $B$-coefficients (see Table~\ref{T6}):
$F = 8a \rightarrow 7a$ and $F = 8b \rightarrow 7b$
with the intermediate values of
$F = J+1$ and the ratio $B_{8a \rightarrow 7a}/B_{8b \rightarrow 7b} = 1.02$.
The second option is shifted by only 5 \ms\ from the
$F = 9a \rightarrow 8a$ transition which is a favoured transition for our case.
Probably, the observed polarization could be explained with this $F = 9a \rightarrow 8a$
hyperfine component as well.
Indeed, taking into account that masing in a spectral line occurs when population is inverted and
the absolute value of the optical depth in the line $\tau > 1$,
emission in the $F = 9a \rightarrow 8a$ component should dominate since $(i)$
the ratio of the Einstein $A$-coefficients
$A_{9a \rightarrow 8a}/A_{8 \rightarrow 7} \approx 2$, and $(ii)$
the intensity of the maser emission is exponentially dependent on $\tau$.
In any case we conclude that the dominance of different hyperfine transitions
in methanol masers should be somehow related to molecular spin alignment within hyperfine structures
when all nuclear spins are of the same sign and the total angular momentum reaches its marginal values.

The orientation of atomic and molecular spins in the interstellar medium was
considered for the first time by Varshalovich (1971) and later on in a series of publications
of different authors (e.g., Burdyuzha \& Varshalovich 1973;
Landolfi \& Landi Degl'Innoceni 1986; Matveenko \etal\ 1988;
Yan \& Lazarian 2007; Zhang \etal\ 2020).
The main physical process behind this phenomenon is, in short, the following.

Atomic and molecular species in isotropic media have randomly oriented spins.
If, however, there is a directed beam of radiation or fast particles then interaction
with the beam will compel the spins to be aligned preferentially in one direction.
In order for spins to be aligned, random collisions with the surrounding gas particles
should not be too effective to flip spins over and thus to randomize their orientation.

In a magnetized medium, each hyperfine level of the methanol molecule with an
angular momentum $F$ will split into $2F + 1$ magnetic sublevels with different energy
depending on the positive or negative Land\'e factor (Lankhaar \etal\ 2018).
Then a difference in populations of hyperfine levels (i.e., the dominance of some of them)
will imply that the spins of the molecules are aligned.
As the projection of the photon spin, $s$,
on its direction of motion is always fully oriented, $s = \pm1$,
whereas the photon state with $s = 0$
is absent due to the transverse character of electromagnetic waves,
the unpolarized beam contains an equal number of photon states with left- and right-hand
circular polarization.
By virtue of this, when a photon is scattered by a molecule,
the spin of the molecule either becomes
aligned anisotropically with an equal number of spins oriented towards or against
the axis of symmetry, or becomes polarized with an unequal number of spins
aligned predominantly in one direction.
And if a magnetic field is present, it will control the orientation of the
plane of polarization. It was shown that a maser with anisotropic pumping
can achieve 100\% polarization (Western \& Watson 1983, 1984; Watson 2009).

We suppose that such polarization of spins can explain the clustering of
the velocity offsets detected in the present work. The involved physical interactions
are still not clear in full detail, but such an analysis which requires extended
calculations is beyond the scope of this paper.

\section{Summary and future prospects}
\label{Sect5}

Looking for the signs of possible non-universal coupling of hypothetical hidden field(s)
to standard matter, we estimate the fractional changes in the electron-to-proton mass ratio, 
$\mu = m_{\rm e}/m_{\rm p}$,
where the mass of the electron is predicted to be directly affected by the Higgs-like coupling and
the mass of the proton is determined mainly by the binding energies of quarks.
The measurements are based on observations of Class~I methanol maser transitions in 
sources distributed in the Milky Way disk 
over the range of the galactocentric distances $4 \la R \la 12$ kpc.
The sources were selected from two surveys performed
by Kim \etal\ (2018) and Yang \etal\ (2020) at
the Korean VLBI Network (KVN) 21-m telescopes in single-dish telescope mode.
Observed were the Class~I $A$-type methanol transitions $7_0 - 6_1$~A$^+$
at 44 GHz, and $8_0 - 7_1$~A$^+$ at 95 GHz.
The value of \dmm\ is measured through the radial velocity offset $\Delta V = V_{44} - V_{95}$
according to Eq.~\ref{Eq3}.

Our main results are as follows.

\begin{itemize}
\item[$\bullet$]
Observations of the 44 GHz line separated by a period of one year reveal
a remarkable stability of the line position with an uncertainty of only $\pm2$ \ms.
\item[$\bullet$]
The measured velocity offsets between the simultaneously measured 44 GHz and 95 GHz lines
are clustered into two groups with the mean values separated by
$\delta_{\scriptscriptstyle \Delta V} = 0.022 \pm 0.003$ \kms\ ($1\sigma$ C.L.).
The presence of two distinguished groups
can be explained if the methanol maser action favors
the following hyperfine transitions:
$F = 5a \rightarrow 4a$ and $6a \rightarrow 5a$ at, respectively, 44 and 95 GHz for the first mode
with the smallest value of $F = J-2$,
and
$F = 9a \rightarrow 8a$ and $10a \rightarrow 9a$ at, respectively, 44 and 95 GHz for the second mode
with the largest value of $F = J+2$.
By this we confirm the suggestion
of Lankhaar \etal\ (2018) that masing involves preferred hyperfine transitions.
The revealed bimodality also confirms 
that the emission from the
44 and 95 GHz transitions arises in the same environment and
is highly cospatial.
\item[$\bullet$]
The measured \dV\ values constrain the
spatial gradient $k_\mu$ of \dmm\ in the Galactic disk
$k_\mu < 2\times10^{-9}$ kpc$^{-1}$
in the range of the galactocentric distances $4 \la R \la 12$ kpc,
while the upper limit on the changes in $\mu$ is \dmm~$<2\times10^{-8}$.
These are the tightest constraints on the spatial $\mu$ variability at present.
\item[$\bullet$]
The rest frequencies of the 44 and 95 GHz methanol transitions
reported in the NIST, NRAO, JPL, and CDMS molecular data bases are given
with underestimated errors.
\end{itemize}

According to these results, future prospects should be the following:
\begin{itemize}
\item[$-$]
The 44 and 95 GHz methanol transitions turned out to be especially
suitable for both \dmm\ estimations and studies of the methanol masing mechanisms.
However, any further development can be possible only if
new laboratory measurements of methanol rest frequencies
with uncertainties of $\sim 1$ kHz will be carried out.
\item[$-$]
Class~I methanol masers have numerous transitions within the millimeter-wavelength range. 
Other Class~I methanol maser pairs, such as the $4_{-1}-3_0$~$E$ line 
at 36 GHz and the $5_{-1}-4_0$~$E$ transition at 84 GHz 
and the series of $J_2-J_1$~$E$ lines near 25 GHz, could also be suitable to estimate \dmm.
\item[$-$]
Observations used in the present work were obtained in course of big surveys and in general were not intended for
high precision measurements of the line positions. 
It is desirable to reobserve with higher S/N and better spectral resolution
at least the selected targets.
\item[$-$]
High spectral resolution polarization measurements
can be also used to obtain quantitative characteristics
of the revealed two groups of Class~I methanol masers.
\item[$-$]
Maser sources being observed with high angular resolution exhibit a complex spatial structure consisting of
multiple spots. That is why interferometric observations of such sources would be of great importance since they
make it possible to control results towards a given target using the measurements of the resolved spots.
\end{itemize}

\section*{Acknowledgements}

S.A.L. and M.G.K were supported by the Russian Science
Foundation under grant No.~19-12-00157.

\section*{Data Availability}

The data underlying this article will be shared on reasonable request to the
corresponding author.

\begin{figure*}
\includegraphics[angle=0,width=1.0\linewidth]{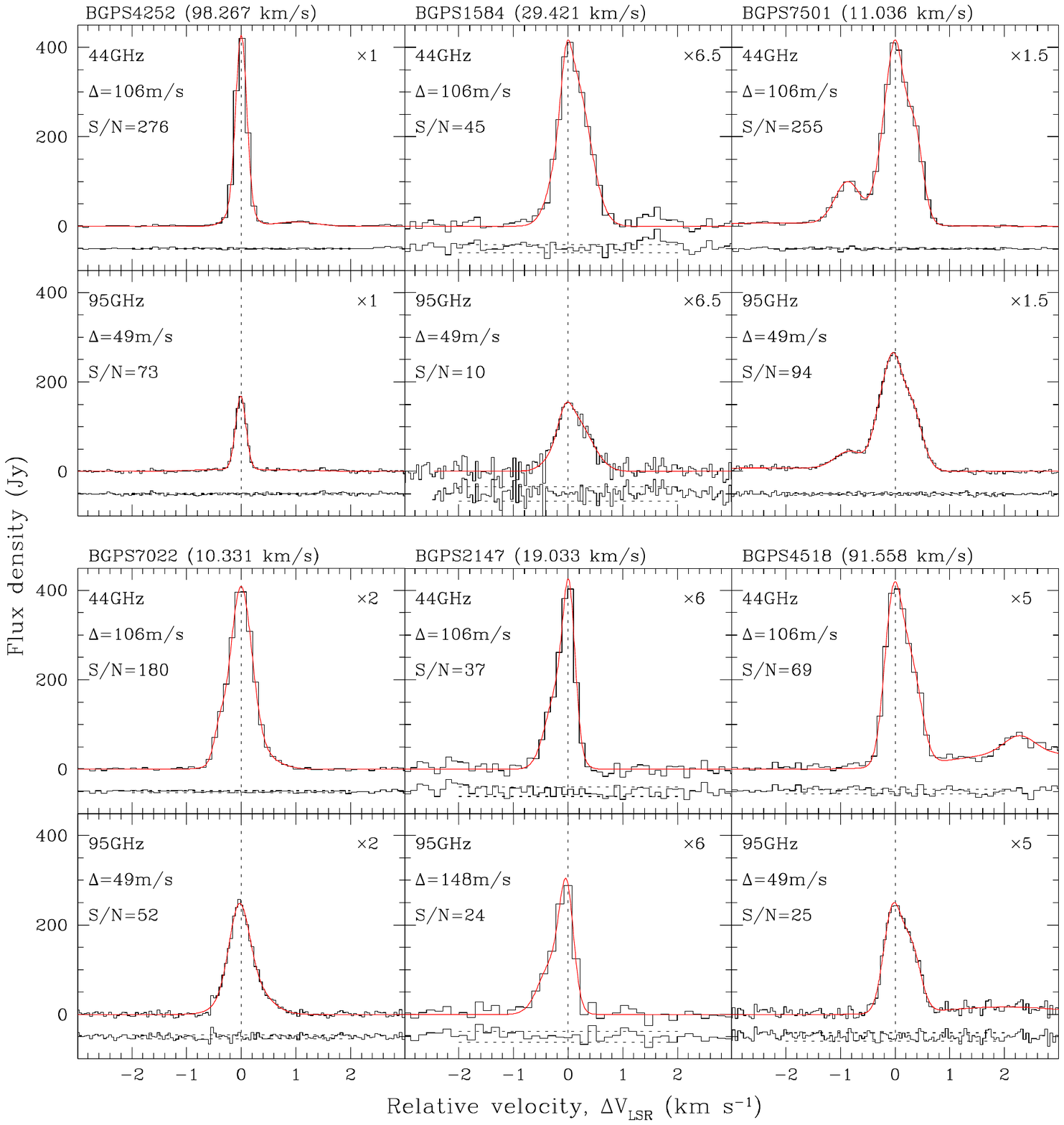}
\vspace{-4.5cm}
\caption{Same as Fig.~\ref{fg3}, but for the BGPS
({\it Bolocam Galactic Plane Survey})  
sources taken from Yang \etal\ (2020) and listed in Table~\ref{T5}.
The BGPS catalogue (version of 1.0.1) can be found at
https://irsa.ipac.caltech.edu/data/BOLOCAM\_GPS/tables/bolocam\_gps\_v1\_0\_1.tbl
}
\label{fg4}
\end{figure*}

\begin{figure*}
\includegraphics[angle=0,width=1.0\linewidth]{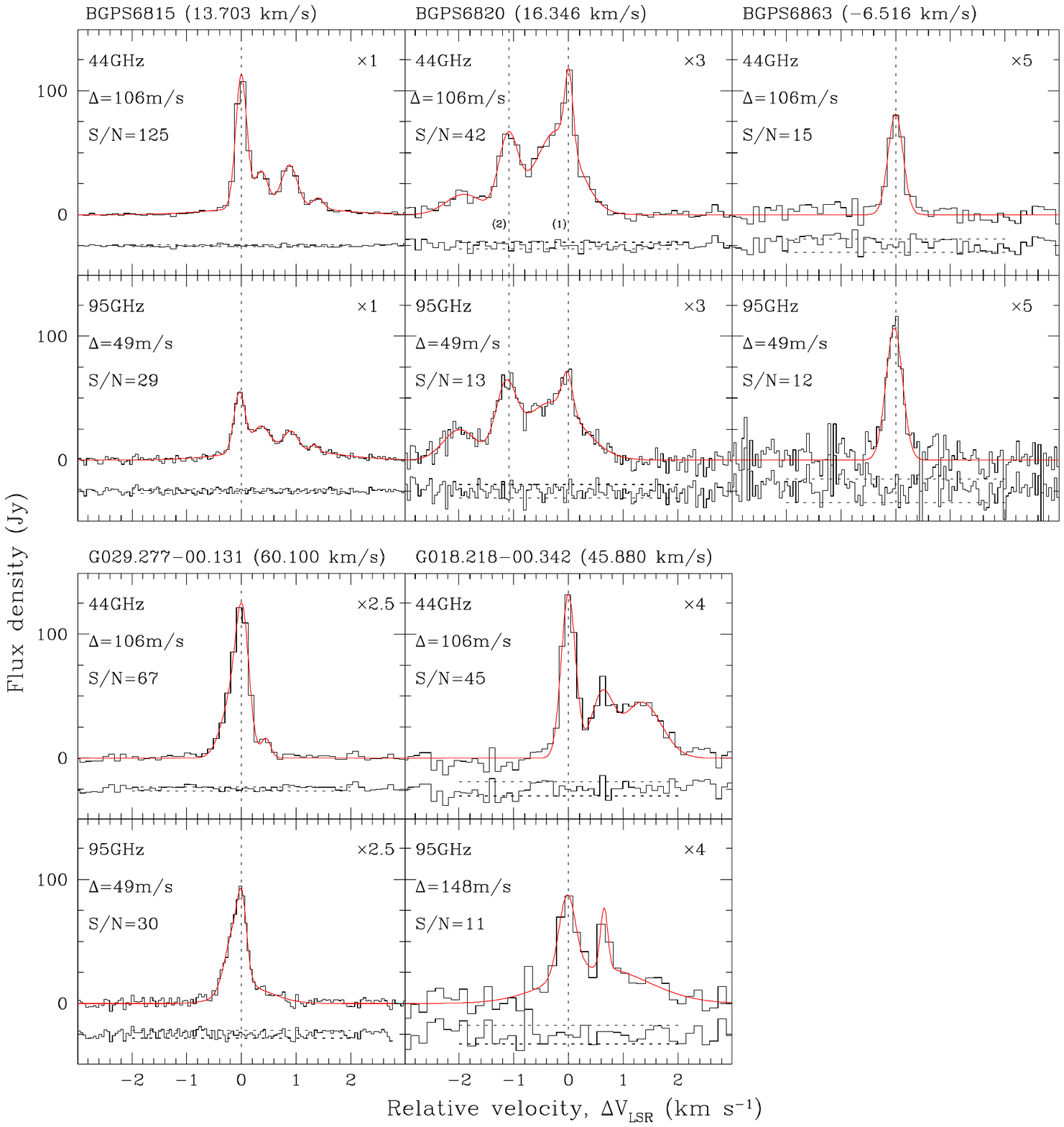}
\vspace{-4.5cm}
\caption{Continuation of Fig.~\ref{fg4} }
\label{fg5}
\end{figure*}

\begin{figure*}
\includegraphics[angle=0,width=1.0\linewidth]{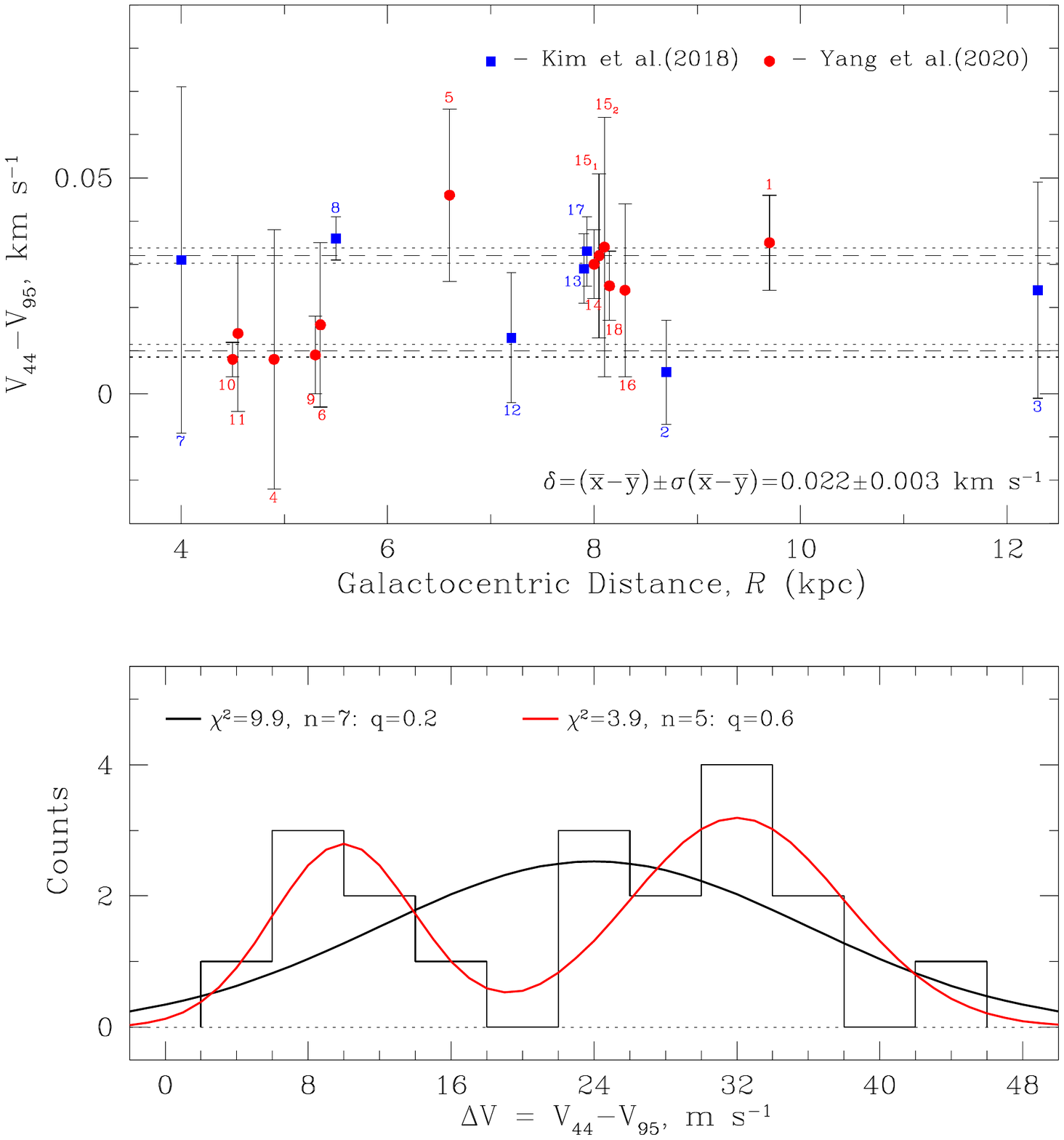}
\vspace{-4.5cm}
\caption{
{\it Upper panel}.
The measured velocity offsets $\Delta V = V_{44}-V_{95}$ against the galactocentric distances
of the Class~I methanol masers
for the rest frequencies of 44069.430 MHz (Pickett \etal\ 1998) and 95169.463 MHz (M\"uller \etal\ 2004).
The blue squares and red dots represent the sources from Tables~\ref{T4} and \ref{T5},
respectively.
The horizontal dashed and dotted lines represent two sample means
$\bar{x} = 0.032$ \kms\ and $\bar{y} = 0.010$ \kms, and their $\pm1\sigma$ boundaries.
Student's $t$-test supports the revealed difference $\delta$ between two groups of masers at $7.3\sigma$.
{\it Lower panel}.
The binned distribution of \dV\ (histogram) compared to the unimodal (black curve)
and bimodal (red curve) distributions.
The corresponding $\chi^2$ values, the numbers of degrees of freedom $n$, and the quantiles
$q$ are shown at the top of the panel.
The probability of the bimodal distribution is 60\% against 20\% for the unimodal distribution
in accord with the $\chi^2$-test for normality.
}
\label{fg6}
\end{figure*}

\begin{figure*}
\includegraphics[angle=0,width=0.5\linewidth]{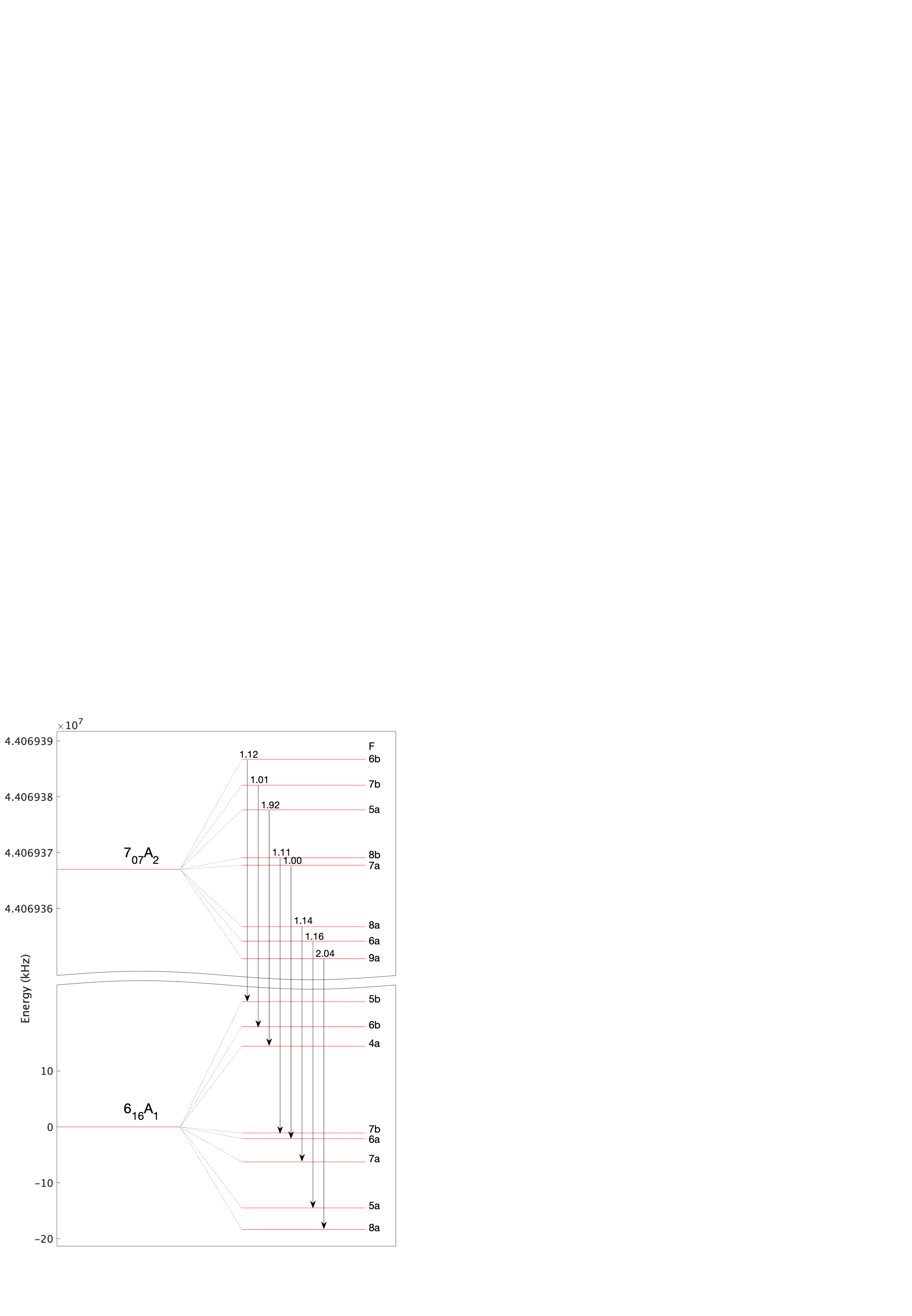}
\vspace{-0.5cm}
\caption{
Hyperfine structure of the torsion-rotation levels of the 44 GHz
($7_{07}A_2 \rightarrow 6_{16}A_1$) transition of methanol 
as calculated in Lankhaar \etal\ (2016).
The energy of the $6_{16}A_1$ torsion-rotation level is set to zero.
The Y-axis is broken in order to show both the hyperfine interactions
($\Delta E \sim 10$ kHz) and the torsion-rotational energy difference
(here $\Delta E \sim 44$ GHz). The torsion-rotation central frequency $f_0 = 44069.367$ MHz
is adopted from Xu \etal\ (2008).
Arrows indicate the strongest hyperfine transitions
with $\Delta F = \Delta J = 1$, with the Einstein $A$ coefficients (in $10^{-7}$ s$^{-1}$)
indicated above (see Table~\ref{T6}).
}
\label{fg7}
\end{figure*}

\begin{figure*}
\includegraphics[angle=0,width=0.5\linewidth]{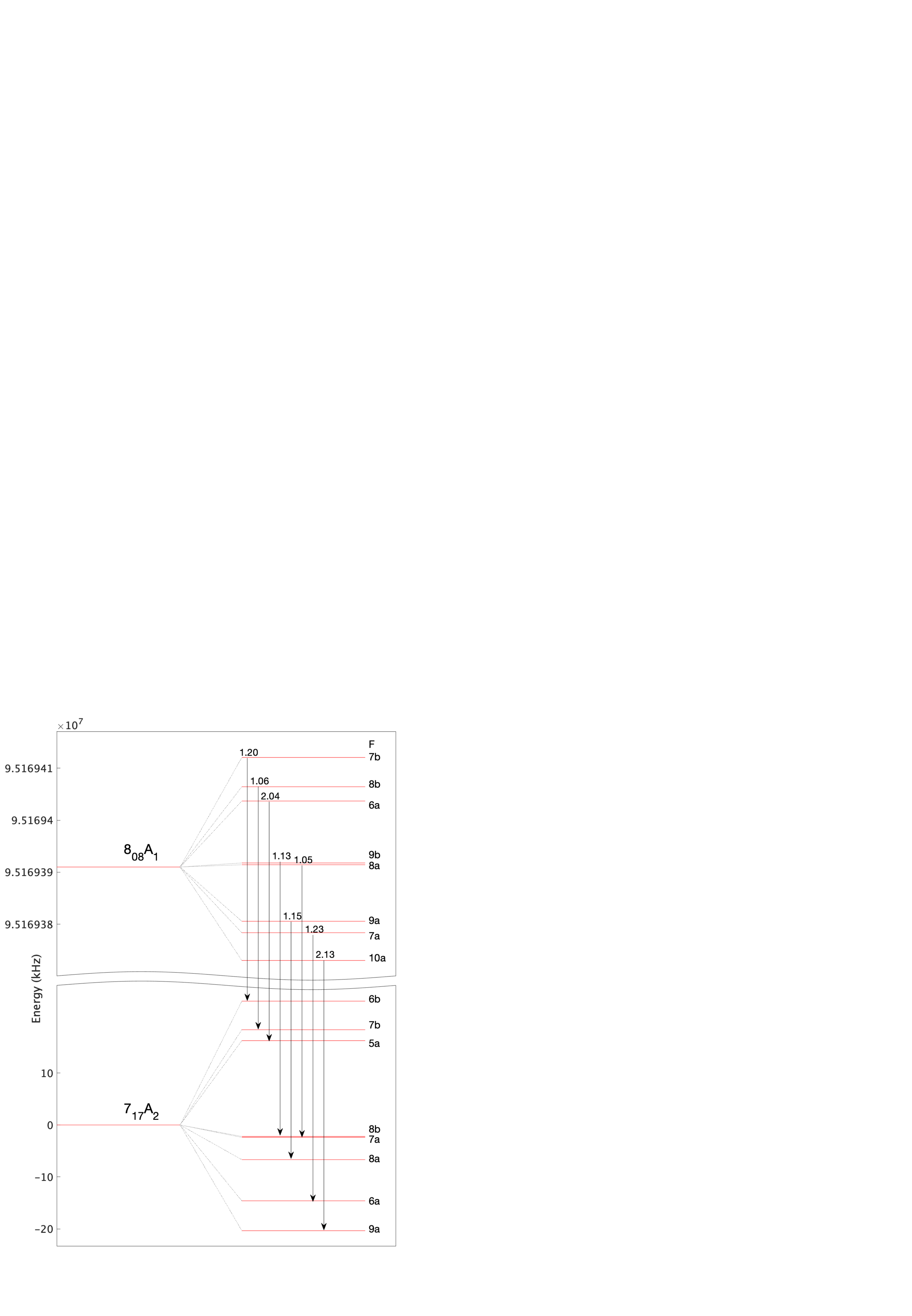}
\vspace{-0.5cm}
\caption{
Hyperfine structure of the torsion-rotation levels of the 95 GHz
($8_{08}A_1 \rightarrow 7_{17}A_2$) transition of methanol 
as calculated in Lankhaar \etal\ (2016).
The energy of the $7_{17}A_2$ torsion-rotation level is set to zero.
The Y-axis is broken in order to show both the hyperfine interactions
($\Delta E \sim 10$ kHz) and the torsion-rotational energy difference
(here $\Delta E \sim 95$ GHz). The central frequency $f_0 = 95169.391$ MHz
is adopted from Xu \etal\ (2008).
Arrows indicate the strongest hyperfine transitions
with $\Delta F = \Delta J = 1$, with the Einstein $A$ coefficients (in $10^{-6}$ s$^{-1}$)
indicated above (see Table~\ref{T7}).
}
\label{fg8}
\end{figure*}

\begin{table*}
\centering
\caption{Selected target sources.
Column 5 gives the radial velocities with respect to the LSR frame, $V_{\rm \scriptscriptstyle LSR}$.
Columns 6 and 7 show the heliocentric, $D$, and galactocentric, $R$, distances. 
Columns 8 and 9 present the peak flux densities, $F_{44}$ and $F_{95}$, at 44 and 95 GHz.
Column 10 shows the bolometric luminosities, $L_{\rm bol}$,
extracted from the RMS catalogue for the sources of Kim \etal\ (2018), 
and from the Hi-Gal catalogue (Elia \etal\ 2021) for the sources of Yang \etal\ (2020) for which
luminosities were updated with respect to the given distances.
}
\label{Tb1}
\begin{tabular}{r l c c r@.l r@.l r@.l r r r}
\hline\\[-5pt]
\multicolumn{1}{c}{No.} & \multicolumn{1}{c}{Source} & R.A. & Dec. 
& \multicolumn{2}{c}{$V_{\rm \scriptscriptstyle LSR}$} &  \multicolumn{2}{c}{$D$}
& \multicolumn{2}{c}{$R$} & \multicolumn{1}{c}{$F_{44}$} & \multicolumn{1}{c}{$F_{95}$} & \multicolumn{1}{c}{$L_{\rm bol}$} \\
 & & (J2000) & (J2000) & \multicolumn{2}{c}{(km~s$^{-1}$)} & \multicolumn{2}{c}{(kpc)} & \multicolumn{2}{c}{(kpc)} &
\multicolumn{1}{c}{(Jy)} & \multicolumn{1}{c}{(Jy)} & \multicolumn{1}{c}{($L_\odot$)} \\
\multicolumn{1}{c}{\tiny (1)} &  \multicolumn{1}{c}{\tiny (2)} & \multicolumn{1}{c}{\tiny (3)}
& \multicolumn{1}{c}{\tiny (4)}
& \multicolumn{2}{c}{\tiny (5)}
& \multicolumn{2}{c}{\tiny (6)}
& \multicolumn{2}{c}{\tiny (7)}
& \multicolumn{1}{c}{\tiny (8)} 
& \multicolumn{1}{c}{\tiny (9)}
& \multicolumn{1}{c}{\tiny (10)} \\ 
\hline\\[-7pt]
1 & BGPS7501$^{\beta}$&06:12:52.90 & +18:00:29.0 & 11&0 & 1&6$^\epsilon$ & 9&7 & 273 & 178 & 4500 \\
2 & RMS149$^{\alpha}$&06:41:10.15 & +09:29:33.6 &  7&2 & 0&6$^\alpha$ & 8&7 & 175 & 55 & 1100 \\
3 & RMS153$^{\alpha}$&06:47:13.36 & +00:26:06.5 &  44&6 & 4&7$^\alpha$ & 12&3 & 25 & 12 & 16000 \\
4 & BGPS1584$^{\beta}$&18:11:59.20 & $-19$:36:03.0 & 29&4 & 3&3$^\gamma$  & 4&9 & 63 & 24 & 260 \\
5 & BGPS2147$^{\beta}$ & 18:20:22.00 & $-16$:14:44.0 & 19&0 & 1&6$^\gamma$ & 6&6 & 67 & 48 & 1400 \\
6 & G018.218-00.342$^{\beta}$&18:25:21.99&$-13$:13:28.5 & 45&9 & 12&3$^\gamma$ & 5&3 & 33 & 22 & 8600 \\
7 & RMS2837$^{\alpha}$ & 18:31:44.08 & $-09$:22:18.5 & 84&0 & 4&9$^\alpha$ & 4&0 & 58 & 48 & 11000 \\
8 & RMS2879$^{\alpha}$ & 18:34:20.89 & $-05$:59:42.5 & 41&8 & 3&0$^\alpha$ & 5&5 & 76 & 52 & 41000 \\
9 & G029.277-00.131$^{\beta}$ &18:45:13.88&$-03$:18:43.9& 60&1 & 3&6$^\delta$ & 5&3 & 49 & 38 & 1300 \\
10 & BGPS4252$^{\beta}$ &18:46:05.60 &$-02$:42:26.0 & 98&3 & 9&0$^\gamma$ & 4&5 & 420 & 168 & 10500 \\
11 & BGPS4518$^{\beta}$ &18:47:41.30 & $-02$:00:21.0 & 91&6 & 5&2$^\gamma$ & 4&5 & 81 & 49 & 49300 \\
12 & RMS3659$^{\alpha}$ & 19:43:11.23 & +23:44:03.6 & 22&5 & 2&2$^\alpha$ & 7&2 & 23 & 17 & 22000 \\
13 & RMS3749$^{\alpha}$ & 20:20:30.60 & +41:21:26.6 & 8&8 & 1&4$^\alpha$ & 7&9 & 65 & 49 & 6500 \\
14 & BGPS6815$^{\beta}$ & 20:35:34.20 & +42:20:13.0  & 13&7 & 1&3$^\delta$ & 8&0 & 107 & 55 & 430 \\
15 & BGPS6820$^{\beta}$ & 20:36:58.10 & +42:11:41.0  & 16&3 & 1&3$^\delta$  & 8&0 & 39 & 24 & 930 \\
16 & BGPS6863$^{\beta}$ & 20:40:28.70 & +41:57:14.0  & $-6$&5 & 3&5$^\gamma$  & 8&3 & 16 & 23 & 2700 \\
17 & RMS3865$^{\alpha}$ & 20:43:28.49 & +42:50:01.8 & 10&3 & 1&4$^\alpha$ & 8&0 & 272 & 161 & 650 \\
18 & BGPS7022$^{\beta}$ & 20:43:28.50 & +42:50:10.0 & 10&3 & 1&1$^\delta$ & 8&0 & 198 & 128 & 570 \\
\hline\\[-5pt]
\multicolumn{13}{l}{\  {\it References:} $^\alpha$Kim \etal\ (2018); $^\beta$Yang \etal\ (2020);
$^\gamma$M\`ege \etal\ (2021); $^\delta$Yang \etal\ (2017); }\\
\multicolumn{13}{l}{$^\epsilon$Rygl \etal\ (2010).}
\end{tabular}
\end{table*}

\begin{table*}
\centering
\caption{List of rest frequencies
of the methanol transitions at 44 GHz and 95 GHz
from different publications.
($m$ -- measured and $c$ -- calculated values).
The uncertainties are shown in parentheses.
}
\label{Tb2}
\begin{tabular}{l r@.l c }
\hline\\[-5pt]
Transition &
\multicolumn{2}{c}{$f$, MHz} & References \\
\hline\\
$7_0 - 6_1$ A$^+$ & 44069&430$^c$     & 1 \\
                  & 44069&410(10)$^m$ & 2 \\
                  & 44069&476(15)$^c$ & 3 \\
                  & 44069&367(10)$^c$ & 4 \\[4pt]
$8_0 - 7_1$ A$^+$ & 95169&475(10)$^m$ & 2  \\
                  & 95169&516(16)$^c$ & 3 \\
                  & 95169&391(11)$^c$ & 4 \\
                  & 95169&463(10)$^m$ & 5 \\
\hline\\[-4pt]
\multicolumn{4}{l}{\  {\it References:} 1~- Pickett \etal\ (1998);}\\
\multicolumn{4}{l}{\  2~- Tsunekawa \etal\ (1995); 3~- Xu \& Lovas (1997);}\\
\multicolumn{4}{l}{\  4~- Xu \etal\ (2008); 5~- M\"uller \etal\ (2004).} \\
\multicolumn{4}{l}{\  {\it Notes:} NIST recommended rest frequencies are based}\\
\multicolumn{4}{l}{\  on [3], whereas NRAO, JPL, and CDMS~-- on [4], see, e.g., }\\
\multicolumn{4}{l}{\  https://splatalogue.online//index.php }
\end{tabular}
\end{table*}

\begin{table*}
\centering
\caption{Comparison of two-epoch observations of Class~I methanol masers.
Measured LSR radial velocities ($V_{44}$)
of the $7_0 - 6_1$ A$^+$ transition at 44069.430 MHz (Pickett \etal\ 1998) and
their differences $\Delta V = V_{2011} - V_{2012}$.
Given in parentheses are statistical errors (1$\sigma$) in the last digits.
}
\label{Tb3}
\begin{tabular}{r c r@.l r@.l }
\hline\\[-3pt]
\multicolumn{1}{c}{RMS} & Date & \multicolumn{2}{c}{$V_{44}$,} &
\multicolumn{2}{c}{$\Delta V = V_{2011}-V_{2012}$,} \\
\multicolumn{1}{c}{ID}  & & \multicolumn{2}{c}{km~s$^{-1}$} &
\multicolumn{2}{c}{km~s$^{-1}$} \\
\hline\\
3841(2)& 2011.04.15 & 0&439(4) &  \multicolumn{2}{c}{} \\
       & 2012.10.12 & 0&437(5) &  $0$&002(6) \\[2pt]
3841(1)& 2011.04.15 & 0&025(6) &  \multicolumn{2}{c}{} \\
       & 2012.10.12 & 0&033(4) & $-0$&008(7) \\
3865& 2011.04.11 & 10&346(5) &  \multicolumn{2}{c}{} \\
    & 2012.10.11 & 10&345(5) & $0$&001(7) \\[2pt]
3749& 2011.04.16 & 8&825(5) &  \multicolumn{2}{c}{}  \\
    & 2012.10.10 & 8&826(4) & $-0$&001(6) \\[2pt]
3659& 2011.04.16 & 22&502(15) &  \multicolumn{2}{c}{} \\
    & 2012.10.11 & 22&494(12) & $0$&008(19) \\[2pt]
2879& 2011.03.23 & 41&780(3) &  \multicolumn{2}{c}{}  \\
    & 2012.05.21 & 41&775(4) & $0$&005(5) \\[2pt]
2837& 2011.03.23 & 83&970(10) &  \multicolumn{2}{c}{}  \\
    & 2012.06.05 & 83&979(9) &  $-0$&009(13) \\[2pt]
2584& 2011.03.21 & 13&210(10) &  \multicolumn{2}{c}{} \\
    & 2012.06.05 & 13&216(9) & $-0$&006(13) \\[2pt]
153 & 2011.03.25 & 44&637(12) &  \multicolumn{2}{c}{} \\
    & 2012.02.15 & 44&643(22) & $-0$&006(25) \\[2pt]
149 & 2011.03.29 & 7&214(7) &  \multicolumn{2}{c}{} \\
    & 2012.03.10 & 7&218(8) & $-0$&004(11)   \\[2pt]

\hline\\[-6pt]
\multicolumn{4}{r}{\  \it mean:} &  $0$&004(8) \\
\hline\\[-4pt]
\end{tabular}
\end{table*}

\begin{table*}
\centering
\caption{Subsample of 7 targets from Kim \etal\ (2018).
Comparison of the velocity offsets $\Delta V = V_{44}-V_{95}$ (in \kms) between
the methanol transitions at 44 and 95 GHz calculated using rest frequencies from
(see Table~\ref{Tb2}):
P~-- Pickett \etal\ (1998); M~-- M\"uller \etal\ (2004);
T~-- Tsunekawa \etal\ (1995); J~-- JPL catalogue; N~-- NIST catalogue.
Indicated are statistical uncertainties ($1\sigma$) in the last digits.
The errors in columns 5--8 are the same as in column 4.
}
\label{T4}
\begin{tabular}{c r@.l r@.l r@.l r@.l r@.l r@.l r@.l }
\hline
 \\[-4pt]
\multicolumn{1}{c}{RMS}
& \multicolumn{2}{c}{ $V^{\rm\scriptscriptstyle P}_{44}$}
& \multicolumn{2}{c}{$V^{\rm\scriptscriptstyle M}_{95}$}
& \multicolumn{2}{c}{$V^{\rm\scriptscriptstyle P}_{44}-V^{\rm\scriptscriptstyle M}_{95}$}
& \multicolumn{2}{c}{$V^{\rm\scriptscriptstyle T}_{44}-V^{\rm\scriptscriptstyle M}_{95}$}
& \multicolumn{2}{c}{$V^{\rm\scriptscriptstyle T}_{44}-V^{\rm\scriptscriptstyle T}_{95}$}
& \multicolumn{2}{c}{$V^{\rm\scriptscriptstyle J}_{44}-V^{\rm\scriptscriptstyle J}_{95}$}
& \multicolumn{2}{c}{$V^{\rm\scriptscriptstyle N}_{44}-V^{\rm\scriptscriptstyle N}_{95}$} \\
\multicolumn{1}{c}{\tiny (1)} &  \multicolumn{2}{c}{\tiny (2)} & \multicolumn{2}{c}{\tiny (3)}
& \multicolumn{2}{c}{\tiny (4)}
& \multicolumn{2}{c}{\tiny (5)}
& \multicolumn{2}{c}{\tiny (6)}
& \multicolumn{2}{c}{\tiny (7)}
& \multicolumn{2}{c}{\tiny (8)} \\
\hline\\
{149} & {7}&{216(7)} & {7}&{211(10)}
&  {$0$}&{005(12)} & {$-0$}&{131}
& {$-0$}&{169}   & {$-0$}&{197} & {0}&{151}
\\
{153} & {44}&{644(6)} & {44}&{620(24)}
& {0}&{024(25)} & {$-0$}&{112}
& {$-0$}&{150} & {$-0$}&{178} & {0}&{170}
\\
{2837}& {83}&{973(7)} & {83}&{942(39)}
& {0}&{031(40)} & {$-0$}&{105}
& {$-0$}&{143} & {$-0$}&{171} & {0}&{177}
\\
{2879}& {41}&{780(2)} & {41}&{744(5)}
& {0}&{036(5)} & {$-0$}&{100}
& {$-0$}&{138} & {$-0$}&{166} & {0}&{182}
\\
{3659}& {22}&{495(5)} & {22}&{482(14)}
& {0}&{013(15)} & {$-0$}&{123}
& {$-0$}&{161} & {$-0$}&{189} & {0}&{159}
\\
{3749}& {8}&{826(3)}  & {8}&{797(7)}
& {0}&{029(8)} & {$-0$}&{107}
& {$-0$}&{145} & {$-0$}&{173} & {0}&{175}
\\
{3865}& {10}&{345(4)} & {10}&{312(7)}
& {0}&{033(8)} & {$-0$}&{103}
& {$-0$}&{141} & {$-0$}&{169} & {0}&{179}
\\
\hline\\[-5pt]
\multicolumn{5}{r}{\it weighted mean:}
& {0}&{030(4)} & {$-0$}&{106}
& {$-0$}&{144} & {$-0$}&{172} & {0}&{176}
\\
\hline\\[-4pt]
\end{tabular}
\end{table*}

\begin{table*}
\centering
\caption{Subsample of 11 targets from Yang \etal\ (2020).
Comparison of the velocity offsets $\Delta V = V_{44}-V_{95}$ (in \kms) between
the methanol transitions at 44 and 95 GHz calculated using rest frequencies from
(see Table~\ref{Tb2}):
P~-- Pickett \etal\ (1998), and M~-- M\"uller \etal\ (2004).
The 1$\sigma$ uncertainties in the last digits are denoted in parentheses.
}
\label{T5}
\begin{tabular}{l r@.l r@.l r@.l }
\hline
 \\[-4pt]
\multicolumn{1}{c}{Source} & \multicolumn{2}{c}{$V^{\rm\scriptscriptstyle P}_{44}$}
& \multicolumn{2}{c}{$V^{\rm\scriptscriptstyle M}_{95}$} &
\multicolumn{2}{c}{$V^{\rm\scriptscriptstyle P}_{44}-V^{\rm\scriptscriptstyle M}_{95} $} \\
\multicolumn{1}{c}{\tiny (1)} &  \multicolumn{2}{c}{\tiny (2)} & \multicolumn{2}{c}{\tiny (3)} &
\multicolumn{2}{c}{\tiny (4)} \\
\hline\\
BGPS4252 & 98&267(3)    & 98&259(2)  & $0$&008(4)    \\
BGPS1584 & 29&421(12)    & 29&413(28) & $0$&008(30)   \\
BGPS7501 & 11&036(3)    & 11&001(11) & $0$&035(11)   \\
BGPS7022 & 10&331(4)    & 10&306(7)  & $0$&025(8)     \\
BGPS2147 & 19&033(17)   & 18&987(10) & $0$&046(20)   \\
BGPS4518 & 91&558(9)    & 91&544(16) & $0$&014(18)     \\
BGPS6815 & 13&703(6)   & 13&673(5)  & $0$&030(8)     \\
BGPS6820(1) & 16&346(5)& 16&314(18) & $0$&032(19)    \\
BGPS6820(2) & 15&260(14)& 15&226(26) & $0$&034(30)    \\
BGPS6863 & $-6$&516(10) & $-6$&540(17) & $0$&024(20)  \\
G029.277--00.131 & 60&100(5) & 60&091(8) & $0$&009(9)    \\
G018.218--00.342 & 45&880(6) & 45&864(18) & $0$&016(19)  \\
\hline\\[-5pt]
\multicolumn{5}{r}{\it weighted mean:} & $0$&017(3)  \\
\hline
\end{tabular}
\end{table*}

\begin{table*}
\centering
\caption{Transitions between hyperfine levels for the strongest components
$\Delta F = \Delta J = 1$ of the 44 GHz methanol maser (Lankhaar \etal\ 2016).
The hyperfine frequency and velocity offsets, $\Delta f$ and $\Delta V$, are given relative to the
central frequency of the corresponding torsion-rotation transition
$f_0 = 44069.367$ MHz adopted from Xu \etal\ (2008).
The Einstein $A$-coefficients are given in the last column.
}
\label{T6}
\begin{tabular}{c c r@.l r@.l c}
\hline
 \\[-4pt]
$F_{\rm up}$ & $F_{\rm down}$
& \multicolumn{2}{c}{$\Delta f$} & \multicolumn{2}{c}{$\Delta V$} & \multicolumn{1}{c}{$A_{ij}$ }\\
 & & \multicolumn{2}{c}{(kHz)} & \multicolumn{2}{c}{(\kms)} & \multicolumn{1}{c}{($10^{-7}$~s$^{-1}$) }\\
\hline\\
5a &   4a &  $-3$&78 & 0&026 &  1.923\\
6a &   5a &     1&64 & $-0$&011 &  1.164\\
6a &   5b & $-35$&26 & 0&240 &  0.798\\
6b &   5a &    34&18 & $-0$&233 &  0.800\\
6b &   5b &  $-2$&73 & 0&019 &  1.125\\
7a &   6a &     2&85 & $-0$&019 &  0.995\\
7a &   6b & $-17$&18 & 0&117 &  0.951\\
7b &   6a &    17&17 & $-0$&117 &  0.957\\
7b &   6b &  $-2$&86 & 0&019 &  1.009\\
8a &   7a &  $-4$&00 & 0&027 &  1.138\\
8a &   7b &  $-9$&18 & 0&062 &  0.877\\
8b &   7a &     8&33 & $-0$&057 &  0.879\\
8b &   7b &     3&16 & $-0$&022 &  1.112\\
9a &   8a &     2&38 & $-0$&016 &  2.037\\
\hline
\end{tabular}
\end{table*}

\begin{table*}
\centering
\caption{Transitions between hyperfine levels for the strongest components
$\Delta F = \Delta J = 1$ of the 95 GHz methanol maser (Lankhaar \etal\ 2016).
The hyperfine frequency and velocity offsets, $\Delta f$ and \dV, are given relative to the
central frequency of the corresponding torsion-rotation transition
$f_0 = 95169.391$ MHz adopted from Xu \etal\ (2008).
The Einstein $A$-coefficients are given in the last column.
}
\label{T7}
\begin{tabular}{c c r@.l r@.l c}
\hline
 \\[-4pt]
$F_{\rm up}$ & $F_{\rm down}$
& \multicolumn{2}{c}{$\Delta f$} & \multicolumn{2}{c}{$\Delta V$} & \multicolumn{1}{c}{$A_{ij}$ }\\
 & & \multicolumn{2}{c}{(kHz)} & \multicolumn{2}{c}{(\kms)} & \multicolumn{1}{c}{($10^{-6}$~s$^{-1}$) }\\
\hline\\
6a &   5a &   $-3$&55  & 0&011 & 2.042\\
7a &   6a &      1&97  & $-0$&006 & 1.227\\
7a &   6b &  $-36$&47  & 0&115 & 0.843\\
7b &   6a &     35&65  & $-0$&112 & 0.845\\
7b &   6b &   $-2$&78  & 0&009 & 1.197\\
8a &   7a &      2&82  & $-0$&009 & 1.052\\
8a &   7b &  $-17$&89  & 0&056 & 1.005\\
8b &   7a &     17&81  & $-0$&056 & 1.009\\
8b &   7b &   $-2$&91  & 0&009 & 1.064\\
9a &   8a &   $-3$&75  & 0&012 & 1.148\\
9a &   8b &   $-8$&22  & 0&026 & 0.962\\
9b &   8a &      7&46  & $-0$&024 & 0.962\\
9b &   8b &      2&99  & $-0$&009 & 1.132\\
10a &  9a &      2&38  & $-0$&008 & 2.133\\
\hline
\end{tabular}
\end{table*}

\bsp    
\label{lastpage}
\end{document}